\documentclass[pdflatex,tex,twocolumn,epjc3]{svjour3}
\RequirePackage[T1]{fontenc}
\usepackage{color}
\newcommand{\beqa}{\begin{eqnarray}}
\newcommand{\eeqa}{\end{eqnarray}}
\usepackage{float}
\usepackage{epsfig}
\usepackage{graphicx}
\usepackage{latexsym}

\usepackage{mathrsfs}
\usepackage{dcolumn}
\usepackage{bm}%

\usepackage{indentfirst}
\usepackage{amssymb,dcolumn}
\usepackage{latexsym}

\usepackage{setspace}
\usepackage{graphicx,float}
\usepackage[colorlinks,linkcolor=blue,citecolor=blue,urlcolor=blue,hyperindex]{hyperref}
\usepackage{color}
\usepackage{slashed}

\journalname{Eur. Phys. J. C}
\begin{document}

\title{Constraints on primordial black holes in the mixed dark matter scenarios using the ratio $\rm (^3{He}+D)/H$}

\author{Yupeng Yang\thanksref{addr1,addr3,e1}, Xiujuan Li\thanksref{addr2} and Gang Li\thanksref{addr1}}
\thankstext{e1}{e-mail: ypyang@aliyun.com}
\thankstext{e2}{e-mail: lxj@qfnu.edu.cn}
\thankstext{e3}{e-mail: gli@qfnu.edu.cn}

\institute{School of Physics and Physical Engineering, Qufu Normal University, Qufu, Shandong, 273165, China \label{addr1}  
\and 
School of Cyber Science and Engineering, Qufu Normal University, Qufu, Shandong, 273165, China \label{addr2}
\and
Joint Center for Particle, Nuclear Physics and Cosmology,  Nanjing University -- Purple Mountain Observatory,  Nanjing, Jiangsu, 
210093, China \label{addr3}}

\date{Received: date / Accepted: date}

\maketitle

\begin{abstract}
We derive the upper limit on the dark matter (DM) fraction in primordial black holes (PBHs) in the mixed DM scenarios. 
In this scenarios, a PBH can accrete weakly interacting massive particles (WIMPs) to form a ultracompact minihalo (UCMH) 
with a density profile of $\rho_{\rm DM}(r)\sim r^{-9/4}$. The energy released from UCMHs due to dark matter annihilation 
has influence on the photodissociation of $^{4}{\rm He}$, producing the $^{3}{\rm He}$ and the D. 
By requiring that the ratio $\rm (^3{He}+D)/H$ caused by UCMHs does not exceed the measured value, 
we derive the upper limit on the dark matter fraction in PBHs. For the 
canonical value of DM thermally averaged annihilation cross section $\left<\sigma v\right>=3\times 10^{-26}\rm cm^{3}s^{-1}$, 
we find that the upper limit is $f_{\rm PBH} < 0.35(0.75)$ for DM mass $m_{\chi}=1(10)~\rm GeV$. 
Compared with other limits obtained by different astronomical measurements, although our limit is not the strongest, 
we provide a different way of constraining the cosmological abundance of PBHs.

\end{abstract}


\section{Introduction} 
\setlength{\parindent}{2em}
Primordial black holes (PBHs) could form via the collapse of large density perturbations exist 
in the radiation dominated universe. It is well known that more massive black hole lives longer ($\tau \propto M_{\rm BH}^{3}$). 
Therefore, taking into account the current age of the Universe, $t\sim \rm 13.7Gyr$, 
PBHs with masses of $M_{\rm PBH}< 10^{15} \mathrm{g}$ have evaporated, while larger PBHs should still exist in the present universe 
\cite{pbhs_review,carr,Carr:2020gox,Carr:2021bzv,Auffinger:2022khh,Choudhury:2023jlt}. 
Recently, LIGO and Virgo detected the gravitational wave signals 
caused by the emerges of black holes~\cite{LIGO}. It has been suggested that some of these signals would be from PBHs, 
which may constitute a fraction of DM depending on their masses
~\cite{Bird:2016dcv,Deng:2021ezy,Wang:2016ana,Choudhury:2023vuj,Choudhury:2023rks,Choudhury:2023hvf,Choudhury:2023kdb,Choudhury:2013woa,Choudhury:2023kam,DAgostino:2022ckg}.

Although DM has been confirmed by many different astronomical observations, its nature remains a mystery~\cite{Choudhury:2015zlc,Choudhury:2015eua}. 
As a popular DM model, weakly interacting massive particles (WIMPs) has been studied widely, and they 
can annihilate into such as photons, electrons and 
positrons (see, e.g., Refs.~\cite{Bertone:2004pz,Jungman:1995df} for a review). 
In the mixed DM scenarios consisting of PBHs and WIMPs, WIMPs can be accreted onto PBHs forming a kind of 
DM structure named ultracompact minihalos (UCMHs) with a density profile $\rho_{\rm DM}(r)\sim r^{-9/4}$~\cite{0908.0735,Eroshenko:2016yve}. 
Since the annihilation rate of WIMPs is proportional to the square of the number density, ${\rm \Gamma} \propto 
n_{\rm DM}^2$, it is expected that the annihilation rate of DM in UCMHs 
is larger than that of classical DM halos. 
\footnote{For a clearer description, we will use DM for referring specifically to WIMPs in the following.}
The gamma-ray flux from UCMHs due to DM annihilation would have contributions to the relevant observations, and those observations 
can be used to investigate the cosmological abundance of PBHs
~\cite{Yang:2020zcu,Yang_2011,Zhang:2021mth,Tashiro:2021xnj,Carr:2020mqm}. 
Moreover, the particles emitted from UCMHs due to DM annihilation 
have interactions with that exist in the Universe, leading to the changes of the thermal history of the Universe. 
The relevant measurements, such as the cosmic microwave background (CMB) and global 21cm signals, 
can be used to investigate these effects and derive the upper limits on the cosmological abundance of PBHs
~\cite{Yang:2020zcu,Tashiro:2021xnj,Yang:2022nlt}. 

It is known that the extra energy injected into the Universe 
during the early epoch can cause the deviation of the CMB from blackbody spectrum
~\cite{carr,McDonald:2000bk,Chluba:2011hw,Chluba:2019nxa,Pani:2013hpa,Nakama:2017xvq,Tashiro:2008sf,PhysRevD.98.023001,McDonald:2000bk,Yang:2022nlt}. 
In the radiation dominated universe, the mass of a UCMH keeps unchanged until the time of matter–radiation 
equality ($z_{\rm eq}\sim 3387$)~\cite{0908.0735}. 
The energy released from UCMHs due to DM annihilation results in the CMB $y$-type distortion, and the upper limits 
on the distortion measured by the Far Infrared Absolute Spectrophotometer experiment 
can be used to constrain the cosmological abundance of PBHs~\cite{Fixsen:1996nj}. In addition to affecting 
the CMB blackbody spectrum, the extra energy injection has influence on the 
photodissociation of $^{4}{\rm He}$ in the early universe~\cite{McDonald:2000bk,Luo:2020dlg}. The observed abundance of 
$^{3}{\rm He}$ and D, produced through dissociating the $^{4}{\rm He}$ nuclei, can be used 
to investigate the cosmological abundance of PBHs. 
Here we will investigate the influence of the energy released from UCMHs due to DM annihilation on the photodissociation of $^{4}{\rm He}$. 
Taking into account the measured upper limit on the ratio $\rm (^3{He}+D)/H<1.10 \times 10^{-4}$~\cite{Copi:1994ev,D-He}, 
we will derive the upper limits on the cosmological abundance of PBHs.

This paper is organized as follows. In Sec. II we briefly review the basic properties of UCMHs, 
and then derive the upper limits on the fraction of DM in PBHs using the measured ratio 
$\rm (^3{He}+D)/H$. The conclusion is given in Sec. III.


\section{The properties of UCMHs and upper limits on the cosmological abundance of PBHs}
\subsection{The basic properties of UCMHs}

It is well known that the cosmological large scale structures are from the density perturbations 
exist in the early universe with a amplitude of 
$\delta \rho/\rho \sim 10^{-5}$. Large density perturbation, e.g., $\delta \rho/\rho > 0.3$, 
would result in the formation of PBHs~\cite{carr}. 
It has been argued that the density perturbation in the range of $10^{-4}<\delta \rho/\rho<0.3$ could result in the formation of 
UCMHs~\cite{0908.0735}. 
However, the simulations have shown that the direct collapse of these large density perturbations is 
not suitable for the formation of UCMHs~\cite{Delos:2017thv,Gosenca:2017ybi}. 
On the other hand, it is found that the UCMHs can be formed through the accretion of DM particles onto PBHs 
after their formation~\cite{Eroshenko:2016yve,Carr:2020mqm,Adamek:2019gns,1985ApJS...58...39B}. 
For our purpose, we will not refer to the detailed formation mechanism of UCMHs and adopt the latter scenario. 

After the formation of PBHs, they can accrete DM particles onto them, resulting in the formation of UCMHs. 
The density profile of DM particles in a UCMH is in the form of~\cite{Tashiro:2021xnj}

\beqa
\rho_{\rm DM }(r,z)=\left\{
\begin{array}{rcl}
\rho_{\rm max}(z)~~~~~~~~~~~~~~~~~~&&{r\le r_{\rm cut}(z)}\\
\rho_{\rm max}(z)\left(\frac{r}{r_{\rm cut}(z)}\right)^{-9/4}&&{r_{\rm cut}(z)< r\le r_{\rm ta}(z_{\rm eq})}\\
\end{array} \right. 
\eeqa
where $\rho_{\rm max}(z)$ is inner core of UCMH taking into account the DM annihilation~\cite{PhysRevD.72.103517}, 

\beqa
\rho_{\rm max} = \frac{m_{\chi}}{\left<\sigma v\right>(t-t_{i})},
\label{eq:rho_max}
\eeqa
where $t_i$ is the formation time of UCMHs. $m_{\chi}$ and $\left<\sigma v\right>$ are the mass and thermally averaged annihilation cross section of DM particles, respectively. $r_{\rm cut}(z)$ is the radius of inner core~\cite{Adamek:2019gns,Tashiro:2021xnj}, 
\beqa
r_{\rm cut}(z)=\left(\frac{\rho_{\rm max}(z)}{\bar \rho_{\rm DM}(z_{\rm eq})}\right)^{-4/9}r_{\rm ta}(z_{\rm eq})
\eeqa
where $\bar \rho_{\rm DM}$ is the mean density of DM, $r_{\rm ta}$ is the turnaround scale of UCMH~\cite{Adamek:2019gns,Tashiro:2021xnj}, 

\beqa
r_{\rm ta}(z) \approx (2GM_{\rm PBH}t^{2}(z))^{1/3}.
\eeqa

Following the work of~\cite{Tashiro:2021xnj}, here we neglect the DM kinetic energy compared with the potential energy, 
therefore, the mass of PBH considered by us is in the range of 
\beqa
M_{\rm PBH}\ge 2.2\times 10^{-4}{\rm M_{\odot}}\left<\sigma v\right>_{26}^{-1/3}m_{\chi,10}^{-73/24},
\label{eq:m_cutoff}
\eeqa
where 
$\left<\sigma v\right>_{26} =\left<\sigma v\right>/10^{-26}{\rm cm^{3}s^{-1}}$ and 
$m_{\chi,10} = m_{\chi}/10\rm GeV$.

\subsection{Upper limits on the cosmological abundance of PBHs}

The annihilation rate of DM particles within a UCMH can be written as~\cite{Zhang:2006fr,Yang:2020zcu},

\beqa
{\rm \Gamma}_{\rm anni}(z)&&=\frac{1}{2}\int n^{2}_{\rm DM}(r,z)\left<\sigma v\right> 4\pi r^{2}dr\nonumber \\
&&=2\pi\frac{\left<\sigma v\right>}{m_{\chi}^2}\int \rho_{\rm DM}^{2}(r,z)r^{2}dr 
\label{eq:anni_rate}
\eeqa 

The energy injection rate of DM annihilation per unit volume can be written as, 

\beqa
{\rm E}_{\rm inj}(z)&&=n_{\rm PBH}(z)2m_{\rm \chi}{\rm \Gamma}_{\rm anni}(z) \nonumber \\
&&=\frac{2m_{\chi}}{M_{\rm PBH}}{\rho_{c,0}}{\rm \Omega}_{\rm PBH}(1+z)^{3}{\rm \Gamma}_{\rm anni}(z)
\eeqa
where $n_{\rm PBH}(z)=\rho_{\rm PBH}(z)/M_{\rm PBH}$ is the number density of PBH at redshift $z$. 
${\rm \Omega}_{\rm PBH}=\rho_{\rm PBH}/\rho_{c,0}$ and $\rho_{c,0}$ is the current critical density of the Universe.

The energy injected into the Universe from UCMHs due to DM annihilation will result in the 
photodissociation of $^{4}{\rm He}$, producing the $^{3}{\rm He}$ or the D. 
Taking this issue into account, the ratio $(^3{\rm He}+\rm D)/H$ can be written as~\cite{McDonald:2000bk},
\beqa
\frac{^3{\rm He}+\rm D}{\rm H}=\int \frac{E_{\rm inj}}{n_{\rm H}}\left[N_{\rm D}(z)+N_{^3\rm He}(z)\right]dt
\label{eq:ratio}
\eeqa
where $N_{\rm D}(z)$ ($N_{^3\rm He}(z)$) is the number of D ($^3{\rm He})$ for a given amount of injected energy at redshift $z$, 
and we will use the results given in Ref.~\cite{Protheroe:1994dt}. For the calculations, we will use the relation of 
$dt=1/H(z)(1+z)dz$, where ${\rm H}(z)={\rm H}_{0}\sqrt{{\rm \Omega}_{\rm m}(1+z)^{3}+
{\rm \Omega}_{{\rm \Lambda}}+{\rm \Omega}_{\gamma}(1+z)^4} $

\begin{figure}
\centering
\includegraphics[width=0.5\textwidth]{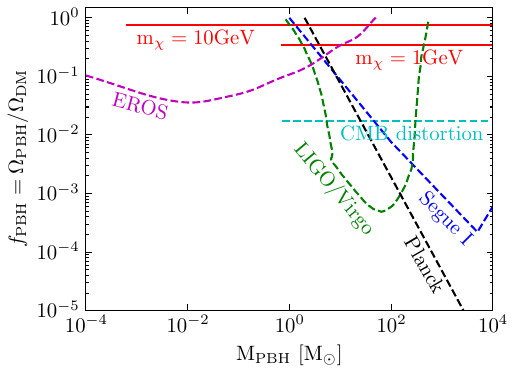}
\caption{Upper limits on the fraction of DM in PBHs, $f_{\rm PBH}={\rm \Omega}_{\rm PBH}/{\rm \Omega}_{\rm DM}$. 
The limits from the ratio $\rm (^3{He}+D)/H$ are shown for the DM mass $m_{\chi}$=1 and 10 GeV (red solid lines). 
The constraints from several other measurements are shown for comparison: 
1) the limits from the CMB $y$-type distortion measured by the FIRAS experiment (labelled ‘CMB distortion')~\cite{Yang:2022nlt};
2) the limits from the estimation on 
the merger rate of PBHs in light of the sensitivity of LIGO/Virgo (labelled ‘LIGO/Virgo')~\cite{Vaskonen:2019jpv}; 
3) the limits from the investigation of the dynamical evolution of stars in the dwarf galaxy Segue I 
(labelled ‘Segue I’)~\cite{segue}; 
4) the constraints from the studies on the anisotropy of CMB caused by the accreting PBHs 
with Planck data (labelled ‘Planck’)~\cite{Poulin:2017bwe}; 
5) the constraints from the studies on the gravitational lensing results measured by EROS (labelled ‘EROS')~\cite{EROS}. 
} 
\label{fig:fraction_compare}
\end{figure}

The ratios $\rm ^3{\rm He}/H$ and $\rm D/H$ have been measured in many celestial 
systems~\cite{Burles:1998zm,Copi:1994ev,Webb:1997mt,carr}. 
Here we use the upper limit on the ratio $\rm (^3{He}+D)/H<1.10 \times 10^{-4}$ for our calculations~\cite{Copi:1994ev}. 
By requiring that the ratio calculated with Eq.~(\ref{eq:ratio}) 
does not exceed the measured value, one can get the upper limits on the cosmological abundance of PBHs. 
The constraints on the fraction of DM in PBHs, $f_{\rm PBH}={\rm \Omega}_{\rm PBH}/{\rm \Omega}_{\rm DM}$, 
are shown in Fig.~\ref{fig:fraction_compare} (red solid lines). 
Here we have set the thermally averaged annihilation cross section of DM as 
$\left<\sigma v\right>=3\times 10^{-26}~\rm cm^{3}s^{-1}$. 
We found that the upper limit is $f_{\rm PBH} < 0.35(0.75)$ for DM mass $m_{\chi}=1(10)~\rm GeV$. 
Since the energy injection rate of DM annihilation (Eq.~(\ref{eq:anni_rate})) is lower for larger DM mass, 
the limit is weaker for larger DM mass. On the other hand, the energy injection rate of DM annihilation per unit volume 
is independent on the PBH mass, ${\rm E}_{\rm inj}\propto n_{\rm PBH}{\rm \Gamma}_{\rm anni}\propto {\rm M}_{\rm PBH}^{-1}{\rm M}_{\rm PBH}$, 
therefore, the limits are independent on the PBH mass for our considerations. 
Note that since we have neglect the DM kinetic energy compared with the potential energy for the formation of UCMHs around PBHs, 
there is a cut off for the upper limit depending on the DM mass (Eq.~(\ref{eq:m_cutoff})). \footnote{A more detailed discussions 
can be found in, e.g, Refs.~\cite{Tashiro:2021xnj,Adamek:2019gns,Boucenna:2017ghj,Cai:2020fnq}. In general, for smaller PBH (smaller than the value of Eq.~(\ref{eq:m_cutoff})), if the kinetic energy of DM is 
not negligible compared to the potential energy during the formation of UCMH, the upper limit on $f_{\rm PBH}$ 
will be weaker than that of larger PBH depending on the mass.}

In the mixed dark matter scenarios, the fraction of DM in PBHs can be constrained 
by many different astronomical 
measurements~\cite{Carr:2020gox,Carr:2021bzv,Carr:2020mqm,Cai:2020fnq,Yang:2020zcu,Tashiro:2021xnj,Boucenna:2017ghj}. 
In Fig.~\ref{fig:fraction_compare}, we also plot several other upper limits on $f_{\rm PBH}$ for comparison.
\footnote{For other constraints one can refer to, e.g., Refs.~\cite{pbhs_review,carr,Carr:2020gox,Carr:2021bzv,Auffinger:2022khh,Cang:2020aoo,yinzhema,Zhou:2021ndx,Dike:2022heo,Zhang:2023rnp}, for a review.} 
The extra energy injected into the early universe can affect the distribution of photons, resulting in the deviation of 
the CMB from blackbody spectrum. For the energy released from UCMHs due to DM annihilation, CMB $y$-type distortion is caused~\cite{Yang:2022nlt} 
and the upper limit on the distortion have been measured by the Far Infrared Absolute Spectrophotometer experiment (FIRAS)~\cite{Fixsen:1996nj}. 
In previous work, we have used the measured results to constrain the fraction of PBHs~\cite{Yang:2022nlt}, 
and the upper limit is shown in Fig.~\ref{fig:fraction_compare} 
for DM mass $m_{\chi}=1\rm GeV$ (cyan dashed line, labelled 'CMB distortion'). 
The upper limit from the ratio $\rm (^3{He}+D)/H$ is weaker by a factor of $\sim 18$ than that from the CMB $y$-type distortion. 
\footnote{Note that here we have used a different density profile of DM in UCMHs compared with that in previous work, 
which can result in the changes of the upper limit slightly.}

 
As mentioned above, the DM annihilation rate in UCMHs is higher than that in classical DM halos. Therefore, 
it is expected that they have significant contributions to 
the extragalactic $\gamma$-ray background (EGB)~\cite{ucmh_scott,Zhang:2021mth,Yang:2020zcu,Yang_2011,Boucenna:2017ghj}. 
The cosmological abundance of PBHs can be constrained using the EGB data measured 
by the Fermi-LAT~\cite{Fermi-LAT:2015bhf,Fermi-LAT:2019yla}. 
The authors of~\cite{Carr:2020mqm} investigated this issue and derived a upper limit of $f_{\rm PBH}<10^{-11}$ 
for $m_{\chi}=10$ GeV. The released energy from UCMHs due to DM annihilation can also inject into the intergalactic medium, 
affecting the thermal history of the Universe and resulting in the changes of the anisotropy of CMB. 
By using the Planck-2018 data, the authors of~\cite{Tashiro:2021xnj} found that the upper limit is $f_{\rm PBH}<3\times 10^{-10}$ 
for $m_{\chi}=10$ GeV. Note that these upper limits are not shown in Fig.~\ref{fig:fraction_compare}, where we have 
set the range of $f_{\rm PBH}$ as $[10^{-5},1]$. 

Note that the constraints on $f_{\rm PBH}$ here depend on the measured upper limits on the ratios $\rm ^3{\rm He}/H$ and $\rm D/H$. 
Compared with the ratio $\rm ^3{\rm He}/H$, $\rm D/H$ is less effected by the cosmological evolution and can be determined well basing on, e.g., 
the observations and analysis of DI and HI lines from damped Lyman-$\alpha$ systems
~\cite{Pitrou:2020etk,Cooke:2017cwo,Kirkman:2003uv,Keith:2020jww}. The ratio $\rm ^3{\rm He}/H$ is less constraining due to the influences of stars 
and few observations in the Galactic disk~\cite{Pitrou:2020etk,2002Natur.415...54B,Jedamzik:2006xz,Geiss,Copi:1994ev,Olive:1996tt}. 
For the future determination of the ratio $\rm D/H$, more damped Lyman-$\alpha$ systems are needed for decreasing the errors, 
which can be achieved by, e.g., the next generation of 30-m class telescopes~\cite{Pitrou:2020etk}. 
The situation is slightly more complicated for the ratio $\rm ^3{\rm He}/H$. 
In the future, an better understanding of stellar nucleosynthesis models and more observations in the
Galactic disk are the key to improve the accuracy of the ratio $\rm ^3{\rm He}/H$~\cite{Keith:2020jww,Pitrou:2020etk}. 

\section{Conclusion}

We have derived the upper limits on the fraction of DM in PBHs in the mixed dark matter scenarios consisting of PBHs and WIMPs. In this scenarios, a PBH accrete WIMPs to form a UCMH with a density profile of $\rho_{\rm DM}(r)\sim r^{-9/4}$. Compared with the classical 
DM halos, the formation time of UCMHs is earlier and the annihilation rate of DM in UCMHs is larger. It is expected that 
the energy released from UCMHs due to DM annihilation has significant influences on different astrophysical processes. 
We have investigated these effects on the ratio $\rm (^3{He}+D)/H$ 
produced through dissociating the $^{4}{\rm He}$ nuclei. By requiring that the ratio caused by the DM annihilation in UCMHs 
does not exceed the measured value of $\rm (^3{He}+D)/H<1.10 \times 10^{-4}$, we obtained the upper limits on the fraction of DM 
in PBHs. We found that the upper limit is $f_{\rm PBH} < 0.35(0.75)$ for DM mass $m_{\chi}=1(10)~\rm GeV$. 
Compared with other limits obtained by different astronomical measurements, although our limit is not the strongest, 
we provide a different way of constraining the cosmological abundance of PBHs. 
On the other hand, the limits on $f_{\rm PBH}$ can be stronger with the accurate measurements of the ratios 
$\rm ^3{\rm He}/H$ and $\rm D/H$, which can be achieved by the future more observations of the related systems 
and better understanding of stellar nucleosynthesis models.

\section{Acknowledgements}
Y. Yang is supported by the Shandong Provincial Natural Science Foundation (Grant No. ZR2021MA021). 
X. Li is supported by the Shandong Provincial Natural Science Foundation (Grant No. ZR2023MA049). 
G. Li is supported by the Taishan Scholar Project of Shandong Province 
(Grant No. tsqn202103062).
\

\bibliographystyle{sn-aps} 

\bibliography{refs}

\end{document}